\documentclass[11pt]{article}
\usepackage{amsmath,amssymb,color}

\makeatletter
\@addtoreset{equation}{section}
\makeatother

\usepackage{amsfonts}

\newcommand{\hoch}[1]{$\, ^{#1}$}

\newcommand{\be}{\begin{equation}}
\newcommand{\ee}{\end{equation}}
\newcommand{\bea}{\setlength\arraycolsep{2pt} \begin{eqnarray}}
\newcommand{\eea}{\end{eqnarray}}
\newcommand{\nn}{\nonumber}

\def\ft#1#2{{\textstyle{\frac{\scriptstyle #1}{\scriptstyle #2} } }}
\def\fft#1#2{{\frac{#1}{#2}}}

\def\0{{\sst{(0)}}}
\def\1{{\sst{(1)}}}
\def\2{{\sst{(2)}}}
\def\3{{\sst{(3)}}}
\def\4{{\sst{(4)}}}
\def\5{{\sst{(5)}}}
\def\6{{\sst{(6)}}}
\def\7{{\sst{(7)}}}
\def\8{{\sst{(8)}}}
\def\sst#1{{\scriptscriptstyle #1}}

\def\cG{{{\cal G}}}
\def\im{{\rm i\,}}

\begin{document}

\begin{flushright}
\hfill{ \
CAS-KITPC/ITP-272\ \ \ \ MIFPA-11-24\ \ \ \ }
\end{flushright}

\vspace{25pt}
\begin{center}
{\large {\bf Conformal Gravity and Extensions of Critical Gravity}}

\vspace{15pt}

H. L\"u\hoch{1,2}, Yi Pang\hoch{3} and C.N. Pope\hoch{4,5}

\vspace{10pt}

\hoch{1}{\it China Economics and Management Academy\\
Central University of Finance and Economics, Beijing 100081, China}

\vspace{10pt}

\hoch{2}{\it Institute for Advanced Study, Shenzhen
University\\ Nanhai Ave 3688, Shenzhen 518060, China}

\vspace{10pt}
\hoch{3} {\it Kavli Institute for Theoretical Physics, Key
Laboratory of Frontiers in Theoretical Physics, Chinese Academy of Sciences, Beijing 100190,
P. R. China}

\vspace{10pt}

\hoch{4} {\it George P. \& Cynthia Woods Mitchell  Institute
for Fundamental Physics and Astronomy,\\
Texas A\&M University, College Station, TX 77843, USA}

\vspace{10pt}

\hoch{5}{\it DAMTP, Centre for Mathematical Sciences,
 Cambridge University,\\  Wilberforce Road, Cambridge CB3 OWA, UK}

\vspace{40pt}

\underline{ABSTRACT}
\end{center}

 Higher-order curvature corrections involving the conformally-invariant
Weyl-squared action have played a role in two recent investigations of
four-dimensional gravity; in critical gravity, where it is added to the
standard cosmological Einstein-Hilbert action with a coefficient tuned to
make the massive ghostlike spin-2 excitations massless, and in a pure Weyl-squared action considered by Maldacena, where the massive spin-2 modes are removed by the imposition of boundary conditions.  We exhibit the connections between the two approaches, and we also generalise critical gravity to a wider class of Weyl-squared modifications to cosmological Einstein gravity where one can eliminate the massive ghostlike spin-2 modes by means of boundary conditions. The cosmological constant plays a crucial role in the discussion, since there is then a ``window'' of negative mass-squared spin-2 modes around AdS$_4$ that are not tachyonic.  We also construct analogous conformal and non-conformal gravities in six dimensions.

\vspace{15pt}

\thispagestyle{empty}





\newpage

\section{Introduction}

Although string theory may provide the most promising candidate for a
quantum theory of gravity, there remains a tantalizing question as to whether
four-dimensional gravity can be quantized in its own right.  A natural
approach, and one of
the simplest, is to consider extending Einstein gravity by adding quadratic
curvature terms, thus rendering the theory perturbatively renormalizable
\cite{stelle1,stelle2}. However, as is typical in a theory with more than
second-order time derivatives, it contains ghostlike modes, in the
form of  massive spin-2 excitations.  There is a way to
 circumvent this problem if one considers three dimensions rather than four,
and so the usual
massless graviton is trivial.  Hence the ghostlike massive modes can
become acceptable upon reversing the sign of the Einstein-Hilbert
action, without in the process creating a ghostlike physical massless
graviton. Examples include the well-studied topologically massive
gravity \cite{tmg1},
and the more recently discovered new massive gravity \cite{nmg}. It was
observed that when a cosmological constant is included, there
exists some critical point \cite{lss} in the parameter space such
that the massive modes disappear and are replaced by modes with logarithmic
coordinate dependence \cite{grujoh}.
The theory can be made ghost free while retaining the standard
sign for the Einstein-Hilbert
action, by truncating out the log modes using standard Brown-Henneaux
AdS$_3$ boundary conditions \cite{bh}.  The theory has subsequently been
generalized to a large class of three-dimensional off-shell
supergravities \cite{abrhst}-\cite{lphybrid}.

   Analogous critical gravities in four dimensions were subsequently
proposed \cite{lpcritical}. The Lagrangian consists of the
Einstein-Hilbert term, a cosmological constant $\Lambda$, and a term
constructed from the square of the Weyl tensor,
with a coupling constant $\ft12\alpha$.\footnote{
Actions with a Weyl-squared term
have also been considered in the context of non-commutative geometry
in \cite{chamconn}.}  It was shown that
there is a critical relationship between $\alpha$ and $\Lambda$ such
that the massive spin-2 modes disappear by coalescing with the massless
modes, resulting again in the appearance of logarithmic
modes \cite{d4log}.  (See also \cite{aflog,ggst}.)  These log modes
are ghostlike in nature \cite{prghost,lllstable}, but their fall-off
behaviour at infinity is slower than the standard massless modes,
and so they can be truncated out by imposing appropriate AdS$_4$
boundary conditions. The resulting theory appears, however, to be somewhat
empty, in that the remaining massless graviton has zero on-shell
energy. Furthermore, the mass and entropy of black holes in the critical
theory both vanish.
This critical phenomenon arises also in higher-dimensional gravities
extended by adding curvature-squared terms \cite{dllpst},  and also
if certain cubic curvature terms are added \cite{sgt}. (See also \cite{liusun} for the $D=3$ case.)

Recently, four-dimensional purely conformal gravity \cite{bohost}, where
there is only a Weyl-squared term, has been
revisited in \cite{maldaconf}.  It was shown that if an
appropriate boundary condition is imposed, then for spherically-symmetric
configurations only the Schwarzschild-AdS metric arises as a black-hole
solution in conformal gravity. Furthermore, its Euclidean action
calculated in conformal gravity modified by a purely topological
contribution from a Gauss-Bonnet term
turns out to match exactly with the action of the same
black hole in Einstein gravity with
a cosmological constant, for an appropriate choice of the coefficient
$\alpha$ of the Weyl-squared term in
conformal gravity.  The black-hole entropy calculated for the
conformal gravity and for the usual Einstein gravity then precisely matches
also.  This leads
to the possibility that the two theories at long wavelengths are
in fact equivalent.

    The Lagrangian for critical gravity, modulo a total derivative that
does not affect the equations of motion, is given by \cite{lpcritical}
\begin{equation}
{\cal L}^{\rm crit}= \sqrt{-g} ( R-2\Lambda +
\ft12 \alpha C^{\mu\nu\rho\sigma} C_{\mu\nu\rho\sigma} ) \,,\label{crit0}
\end{equation}
where $C_{\mu\nu\rho\sigma}$ is the Weyl tensor.  It turns out
that the value for $\alpha$ required for criticality, namely
$\alpha=3/(2\Lambda)$, is of precisely the same magnitude as that for the
Weyl-squared coupling coefficient obtained in
\cite{maldaconf} for conformal gravity by imposing the
Euclidean action matching condition
described above.
Thus the essentially vacuous nature of critical gravity is a reflection of
the equivalence of the cosmological Einstein-Hilbert
and the conformal theories.

    In section 2, we review both
critical gravity and the Einstein/Conformal Gravity duality
conjecture. In conformal gravity, there exist ghostlike massive spin-2
modes in the AdS$_4$ background,
satisfying $(\square -\ft23\Lambda - M^2) h_{\mu\nu}=0$, in
addition to the massless spin-2 modes satisfying
$(\square -\ft23\Lambda) h_{\mu\nu}=0$.  Spin-2 representations in AdS$_4$ are
characterised by their lowest-energy $E_0$, which is given by
\be
E_0= \fft32 \pm \sqrt{\fft94 - \fft{3}{\Lambda}\, M^2}\,.\label{E0}
\ee
The representation is unitary if $E_0\ge 3$, and hence $M^2\ge 0$ \cite{kkrep}.  The
time-dependence of the modes is proportional to $e^{-\im E_0 t}$, and
so by analogy with the situation in Minkowski spacetime,
modes may be defined to be tachyonic if $E_0$ becomes complex, thus
leading to exponential growth in time.  From (\ref{E0}), the absence of
tachyons therefore requires\footnote{For scalar fields, the analogous
requirement that $E_0$ be real is equivalent to the Breitenlohner-Freedman
bound \cite{breifree}.}
\be
M^2 \ge \fft34 \Lambda\,.\label{tachyon}
\ee
Interestingly, although the massive modes in conformal gravity
have $M^2<0$, the ``mass''-squared is
not sufficiently negative to violate the bound (\ref{tachyon}),
and so although they are not unitary representations they are not tachyonic.
 However, the radial fall-off of these modes is slower than that for
modes with $M^2\ge 0$. In fact they fall off more slowly even than the
logarithmic
modes. Thus these non-unitary modes can be truncated out by imposing
appropriate
boundary conditions, leaving only the massless graviton.  The
vanishing on-shell energy of the massless graviton in critical
gravity implies that its energy in conformal gravity is exactly the same
as it is in cosmological
Einstein gravity.

   In section 3, we obtain new unitary four-dimensional gravities,
by generalising the parameter choices made for critical gravity
in \cite{lpcritical}.  For critical gravity, the unitarity
requirement $M^2\ge 0$ was imposed for the
spin-2 modes.  However, as noted above,
the absence of tachyonic modes in $D=4$ is less
restrictive than this, and $M^2$ can be negative provided that (\ref{tachyon})
is still satisfied.
This implies we can
choose the coupling $\alpha$ for the Weyl-squared term in (\ref{crit0}) so
that the massive spin-2 modes have $ 3\Lambda /4\le M^2<0$.  These ghostlike
modes are classically stable, but can be truncated out by imposing appropriate
boundary conditions, just as was done for conformal gravity in
\cite{maldaconf},
leaving only the unitary massless graviton modes.  Within this broader
class of cosmological gravity plus Weyl-squared theories,
critical gravity's specific problem of becoming vacuous after truncating the
ghostlike modes is circumvented. Furthermore, since the broader
class of theories has a range of allowable values  for the parameter
$\alpha$, rather than a single critical choice,
the possibility of finding a stable fixed point under the
renormalization group flow becomes less demanding.

        In section 4, we generalize these results to six dimensions.
There are three conformally-invariant structures in $D=6$.  Two of these are
the two independent invariants built from the cube of the
Weyl tensor.  The third is essentially built from second derivatives
of the square of the Riemann curvature.  In order to obtain a
conformal equivalence to Einstein gravity, it is in particular necessary that
Einstein metrics should also be solutions of the conformal gravity.
Indeed, it was already observed that there exists a specific linear combination
of the three conformal structure such that Riemann curvature squared
and cubed terms all vanish \cite{hs}.  As in $D=4$, we find that the
conditions on the coefficients required for critical gravity are
exactly the same as those implied by requiring
Einstein/conformal gravity duality.  We then observe that we can again
construct a more general family of six-dimensional gravities for which the
massive spin-2 modes can be eliminated by boundary conditions.

   The paper ends with conclusions in section 5.  In an appendix, we collect
some of the detailed calculations for the six-dimensional theories.

\section{Critical vs Conformal Gravity in Four Dimensions}

     The Lagrangian of four-dimensional critical gravity discussed in
\cite{lpcritical},
\begin{equation}
{\cal L} = {\cal L}_0 + {\cal L}_1\,,\label{genlag}
\end{equation}
contains two parts.  The first is the usual Einstein-Hilbert
term with a cosmological constant,
\begin{equation}
{\cal L}_0 =\sqrt{-g} (R - 2\Lambda)\,.\label{lag0}
\end{equation}
The second term is quadratic in curvature, namely the square of the Weyl tensor
together with a Gauss-Bonnet term
which is a total derivative:
\begin{eqnarray}
{\cal L}_1 &=& - \ft13\alpha \sqrt{-g}(R^2 - 3 R^{\mu\nu}
R_{\mu\nu})\cr
&=& \ft12 \alpha \sqrt{-g}(C^{\mu\nu\rho\sigma}C_{\mu\nu\rho\sigma} -E_4)\,,
\end{eqnarray}
where
\be
E_4 = R^{\mu\nu\rho\sigma} R_{\mu\nu\rho\sigma} -4 R^{\mu\nu} R_{\mu\nu} + R^2
\ee
is the Gauss-Bonnet invariant whose integral is proportional to the
Euler number.  Being a total derivative in four dimensions, $E_4$ does not
contribute to the equations of motion.

   The Lagrangian ${\cal L}_1$ is proportional to the one for
conformal gravity discussed in \cite{bohost}.
Defining ${\cal L}^{\rm
conf}(\alpha)\equiv -{\cal L}_1$, we have
\begin{equation}
{\cal L}={\cal L}_0 - {\cal L}^{\rm conf}(\alpha)\,.
\end{equation}
The Lagrangian admits Einstein metrics as solutions, with cosmological constant
equal to $\Lambda$. Included amongst these is the AdS$_4$ vacuum solution,
whose curvature is given by
\begin{equation}
R_{\mu\nu}=\Lambda\, g_{\mu\nu}\,,\qquad R= 4\Lambda\,,\qquad
R_{\mu\nu\rho\sigma}= \fft{\Lambda}{3}\, (g_{\mu\rho} g_{\nu\sigma}
 -g_{\mu\sigma} g_{\nu\rho})\,.\label{AdS4}
\end{equation}
Writing the varied metric as $g_{\mu\nu}\rightarrow g_{\mu\nu} +
h_{\mu\nu}$, and so $\delta g_{\mu\nu}=h_{\mu\nu}$, the linearized
equations of motion were given in \cite{lpcritical}.  Choosing
the gauge condition
\begin{equation}
\nabla^\mu h_{\mu\nu}= \nabla_\nu h\,,\label{gauge}
\end{equation}
it was shown that trace part $h$ vanishes by virtue of the equations of motion.
The transverse and traceless spin-2 modes satisfy
\begin{equation}
-\alpha (\Box - \ft23\Lambda) (\Box - \ft23\Lambda - M^2)
h_{\mu\nu}=0\,,\label{box2}
\end{equation}
where
\begin{equation}
M^2=\ft23\Lambda - \fft1{\alpha}\,.\label{genmass}
\end{equation}

   The spectrum contains massless graviton modes
$h_{\mu\nu}^{(m)}$ and also massive spin-2 modes $h_{\mu\nu}^{(M)}$.
Their on-shell energies are given by \cite{lpcritical}
\begin{eqnarray}
E_{\rm massless} &=& -\fft{1}{2\kappa^2T}\, (1-\ft23\alpha\Lambda)\,
   \int\sqrt{-g}\, d^4x\, \nabla^0 h^{\mu\nu}_{(m)}\,
   \dot h^{(m)}_{\mu\nu}\,,
\label{masslessE}\\
E_{\rm massive} &=& \fft{1}{2\kappa^2T}\,(1-\ft23\alpha\Lambda)\,
   \int\sqrt{-g}\, d^4x\, \nabla^0 h_{(M)}^{\mu\nu}\,
   \dot h^{(M)}_{\mu\nu}\,,
\label{massiveE}
\end{eqnarray}
where the integration over the time coordinate is taken over an interval
$T$, which one could take to be the natural time periodicity of AdS$_4$, or
else just send it to infinity.
Since the integrals themselves both give negative quantities,
it follows that ghost modes are unavoidable in general.  In
\cite{lpcritical}, the parameter $\alpha$ was chosen to have the critical
value given by
\begin{equation}
\alpha = \alpha^{\rm crit}\equiv
\fft{3}{2\Lambda}\,,\label{alphacritical}
\end{equation}
implying that $M^2=0$.  In this case, because the massive modes coalesce
with the massless ones, one obtains new solutions to (\ref{box2}) that
are annihilated by neither second-order factor.  These modes have logarithmic
dependence on the AdS$_4$ radial coordinate, and they can be truncated out
by imposing an appropriate AdS boundary
condition. The resulting critical gravity is however rendered
essentially trivial, since the energy (\ref{masslessE}) for
the surviving massless mode
vanishes.  Furthermore, the mass and the entropy of the
Schwarzschild black hole vanish also. The mass formulae for black
holes in extended gravity can be found in
\cite{destek0,destek,pang}.

      In a new development in higher-derivative gravity, four
dimensional conformal gravity \cite{bohost} was revisited in
\cite{maldaconf}. It was observed that the Euclidean action for the
Schwarzschild AdS black hole computed from ${\cal L}^{\rm conf}$ is
identical to that calculated from the pure cosmological Einstein-Hilbert
Lagrangian ${\cal L}_0$, provided that the parameter
$\alpha$ is chosen to take the critical value given in
(\ref{alphacritical}).  The black hole entropy matches also.
We have checked that the actions also match for the Kerr-AdS black hole.
It was proposed in \cite{maldaconf} that subject to the imposition
of appropriate boundary conditions, the
Lagrangians ${\cal L}_0$ and ${\cal L}^{\rm conf}$ are equivalent at
the critical point, in the long wavelength regime. From this point
of view, the ``triviality'' of critical gravity can be easily
understood, since critical gravity (\ref{genlag}) is given by
\begin{equation}
{\cal L}^{\rm crit} ={\cal L}_0 - {\cal L}^{\rm conf}(\alpha^{\rm
crit})\,,
\end{equation}
and so one is subtracting two Lagrangians that describe the same IR physics.
The vanishing in critical gravity of the graviton energy and also
the black hole mass and
entropy further establish the equivalence of ${\cal L}_0$ and ${\cal
L}^{\rm conf}$ at the critical point.

   It should be remarked that there is an issue of ghost modes in
conformal gravity.  The linearized equation of motion following from ${\cal
L}^{\rm conf}$ is given by
\begin{equation}
\alpha (\Box - \ft23\Lambda) (\Box - \ft43\Lambda) h_{\mu\nu}=0\,.
\end{equation}
This implies that
\begin{equation}
M^2=\ft23 \Lambda \,,\label{massinconf}
\end{equation}
which is negative
since $\Lambda <0$ for AdS$_4$.  The energies of the on-shell massless and
massive spin-2 modes are given by
\begin{eqnarray}
E^{\rm conf}_{\rm massless} &=& -\fft{\alpha \Lambda}{\kappa^2T}
   \int\sqrt{-g}\, d^4x\, \nabla^0 h^{\mu\nu}_{(m)}\,
   \dot h^{(m)}_{\mu\nu}\,,
\label{masslessE1}\\
E^{\rm conf}_{\rm massive} &=& \fft{\alpha\Lambda}{\kappa^2T}
   \int\sqrt{-g}\, d^4x\, \nabla^0 h_{(M)}^{\mu\nu}\,
   \dot h^{(M)}_{\mu\nu}\,.
\label{massiveE1}
\end{eqnarray}
Thus we see that $\alpha$ has to be negative for the massless
graviton to have positive energy; meanwhile, the massive graviton
has negative energy.

   The ``mass'' squared, $M^2$, of the massive graviton
(\ref{massinconf}) in conformal gravity is negative, suggesting the
possibility that these modes might be tachyonic.  As we mentioned in the
introduction, the $SO(2,3)$ representations for massive spin-2 modes in
AdS$_4$ are characterised by their lowest energy $E_0$, which is given in
terms of $M^2$ by (\ref{E0}).
   From now on, we shall for convenience take
\be
\Lambda =-3\,,
\ee
so that the AdS$_4$
has ``unit radius.''  The reality of $E_0=\ft32\pm \sqrt{\ft94+M^2}$
therefore requires that
\begin{equation}
M^2 \ge M^2_{\rm min}\equiv - \fft94\,.\label{tachyon2}
\end{equation}
As can be seen from the explicit expressions for the massive
spin-2 modes obtained in \cite{d4log}, which have time dependence
of the form $e^{-\im\, E_0 t}$, the condition that $E_0$ be real ensures
that the modes do not grow exponentially in time.  This is essentially the
statement of the absence of tachyons.
The massive spin-2 modes in conformal gravity, which have $M^2=-2$,
lie within the bound (\ref{tachyon2}), and so they are not tachyonic.

   The radial dependence of the modes with $M^2_{\rm min}\le M^2<0$
exhibits a slower fall off at large distance
than that for modes with $M^2\ge 0$.
In fact they fall off more slowly even than the log modes.  They
can therefore
be truncated out by imposing an appropriate asymptotic
boundary condition, as was described in \cite{maldaconf}. This is essentially
the same boundary condition that can be used to truncate out the logarithmic
 modes in critical gravity.

\section{New Unitary Gravities in Four Dimensions}

  After the truncation of the massive modes, the conformal gravity
described by
${\cal L}^{\rm conf}(\alpha_{\rm crit})$ can be viewed as being equivalent,
at the classical
level, to cosmological Einstein gravity ${\cal L}_{\rm 0}$ \cite{maldaconf}.
It should, however, be emphasised that conformal gravity admits
Einstein metrics with arbitrary cosmological constant as solutions, and
so for a given value of $\alpha$ the
equivalence to Einstein gravity holds only for the specific value
$\Lambda=3/(2\alpha)$ appearing in ${\cal L}_0$.

   It is natural to consider the more restrictive case where the theory
has a unique scale for the AdS vacuum determined by the cosmological constant in the theory. We then need to consider
the Lagrangian (\ref{genlag}).  As discussed earlier, the mass of
the massive spin-2 modes in this theory is given by (\ref{genmass}).
In \cite{lpcritical}, it was required that $M^2\ge 0$, so that these
modes will correspond to unitary representations of $SO(2,3)$.  For
$M^2>0$, they fall off faster than the massless
modes, and so they could not be truncated out by imposing boundary
conditions at infinity.  Thus one would be stuck with having non-truncatable
ghostlike modes, except in the critical case where  $M=0$,
for which the resulting logarithmic modes can be truncated out on account
of their slower fall off.

   An
alternative choice is to choose the $\alpha$ parameter so that $M^2$
lies in the range
\begin{equation}
-\fft94 \le M^2 <0\,.\label{d4notachyon}
\end{equation}
Within this range, the massive modes are non-tachyonic
and classically stable in the sense
that there is no exponential growth in the time.  Since, however, they fall
off more slowly than those with $M^2\ge 0$, one can impose
boundary conditions that eliminate them from the spectrum whilst retaining
the massless modes.\footnote{Note that the $E_0=0$ branch of the
massless solution from
(\ref{E0}) is truncated out for the same reason.}  The condition
(\ref{d4notachyon}) is satisfied by either $\alpha \ge 4$ or $\alpha
< -\ft12$.  It follows from (\ref{masslessE}) that the choice
$\alpha <-\ft12$ implies that the massless graviton has negative
energy.  On the other hand for the choice of $\alpha \ge 4$, the
energy of the massless graviton remains positive.  Of course, in
this case, the massive modes would have negative energy. However, as
we discussed, these modes can be eliminated by imposing appropriate
boundary conditions, leaving just the non-trivial positive-energy
massless graviton.

   As we have seen, by allowing the possibility of having non-tachyonic
but negative-$M^2$ massive modes, which can then be eliminated by boundary
conditions, we have now arrived at a 1-parameter family ($\alpha\ge4$) of
extended gravity theories that describe just unitary massless spin-2
fields.  At the quantum level, having such a family broadens the
chances for finding an ultra-violet fixed point of the renormalization
group flow that lands within the class of acceptable theories.  This
may improve the prospects for obtaining a consistent theory of
quantum gravity.

\section{Generalisations to Six Dimensions}

We now turn our attention to six dimensions. Conformal gravities in
$D=6$ have been previously studied (see, for example,
\cite{bpb,ds,erd}). Three independent structures can arise in the
Lagrangian.  Their explicit forms are (see \cite{bcn}, and also
\cite{mets,oliray})
\bea
I_1 &=& C_{\mu\rho\sigma\nu} C^{\mu\alpha\beta\nu}
 C_{\alpha}{}^{\rho\sigma}{}_\beta\,,\nn\\
I_2 &=& C_{\mu\nu\rho\sigma} C^{\rho\sigma\alpha\beta}
           C_{\alpha\beta}{}^{\mu\nu}\,,\nn\\
I_3 &=& C_{\mu\rho\sigma\lambda}\Big(\delta^\mu_\nu\, \Box +
    4R^\mu{}_\nu - \fft65 R\, \delta^\mu_\nu\Big) C^{\nu\rho\sigma\lambda}
   + \nabla_\mu J^\mu\,,
\eea
where $\nabla_\mu J^\mu$, which does not contribute to the equations of
motion, can be found in \cite{bcn}.
In general,
a Lagrangian of the form $\sqrt{-g}\, c_i I_i$
will give equations of motion that are not satisfied by arbitrary Einstein
metrics.  However, for a specific choice of the $c_i$ (unique up to overall
scaling), the equations of motion will be satisfied by any Einstein metric.
This same linear combination has the feature that, modulo total derivatives,
all terms of
cubic and quadratic order in the Riemann tensor are absent \cite{hs}.
In this form, the Lagrangian is given by
\begin{eqnarray}
e^{-1} {\cal L}_{\rm conf} &=&\beta ( 4 I_1 + I_2 - \ft13 \widetilde I_3 -
\ft1{24} E_6 + \nabla_\mu \widetilde J^\mu)\cr &=&\beta
\Big(RR^{\mu\nu}R_{\mu\nu}-\ft{3}{25} R^3-2R^{\mu\nu}R^{\rho\sigma}
R_{\mu\rho\nu\sigma} -R^{\mu\nu}\Box R_{\mu\nu}+\ft{3}{10}R\Box
R\Big) \,,\label{d6conflag}
\end{eqnarray}
where $\widetilde I_3= I_3-\nabla_\mu J^\mu$, and
the total derivative $\nabla_\mu \widetilde J^\mu$ can
be derived from \cite{hs}. Note that $E_6$ is the Euler
integrand, given by
\begin{equation}
E_6= \epsilon_{\mu_1\nu_1 \mu_2\nu_2 \mu_3\nu_3} \epsilon^{\rho_1
\sigma_1 \rho_2 \sigma_2 \rho_3\sigma_3} R^{\mu_1
\nu_1}{}_{\rho_1\sigma_1} R^{\mu_2 \nu_2}{}_{\rho_2\sigma_2}
R^{\mu_3 \nu_3}{}_{\rho_3\sigma_3}\,.
\end{equation}

   Any Einstein metric (or in fact any metric conformal to Einstein
metric) will be a solution to the theory following from (\ref{d6conflag}).
In particular we may consider the Schwarzschild-AdS black hole,
satisfying $R_{\mu\nu}=-5 g_{\mu\nu}$, with the
metric
\begin{eqnarray}
ds^2 &=& - f dt^2 + f^{-1}\,dr^2  + r^2 d\Omega_4^2\,,\cr 
f&=& 1 + r^2 - \fft{\mu}{r^3}\,.\label{d6bh}
\end{eqnarray}
This is also a solution to Einstein gravity with a
cosmological constant, described by the Lagrangian
\begin{equation}
e^{-1}{\cal L}_0 = R + 20\,.\label{d6einlag}
\end{equation}
Note that we have chosen the cosmological constant so that the
AdS$_6$ vacuum is of unit radius.  The thermodynamic
quantities for the  black hole (\ref{d6bh}) are given by
\begin{equation}
T=\fft{3 + 5r_+^2}{4\pi r_+}\,,\qquad S=\ft23\pi^2 r_+^4\,,\qquad
M=\ft23 \pi r_+^3(1 + r_+^2)\,,
\end{equation}
where $r_+$ is the horizon radius.
The Euclidean action is given by
\begin{equation}
I_6^{\rm Ein} = \fft{2\pi^2 r_+^4 (1-r_+^2)}{3(3+5r_+^2)}\,.
\end{equation}

   Substituting the Euclideanised solution (\ref{d6bh}) into
the action $I_6^{\rm conf}=\int d^6x {\cal L}^{\rm conf}$, we find
that the contribution from $(4 I_1 + I_2 - \ft13 \widetilde I_3)$ converges.
The contribution from the $\nabla_\mu \widetilde J^\mu$ term vanishes.
The integral
of $E_6$ itself diverges, but if, following the same strategy as in
\cite{maldaconf}, one adds in the associated boundary term that arises
in the definition of the Euler number for manifolds with boundary,
it contributes a pure topological number.  The Euclidean action
$I_6^{\rm conf}$ then turns out to be
proportional to $I_6^{\rm Ein}$.  To be specific, we have
\begin{equation}
I_6^{\rm Ein} = I_6^{\rm conf} \Big|_{\rm \beta = -\fft{1}{24}}\,.
\end{equation}
We have also checked this equality for the Kerr-AdS black hole
\cite{glpp1,glpp2}, using also results from \cite{gipepo}, in the
case that the two angular momenta are equal.
It is straightforward to check, using the Wald formula \cite{Iyer}, that the
Schwarzschild-AdS
black hole entropy matches exactly also.  This suggests, therefore,
that as in $D=4$, Einstein gravity emerges from conformal gravity.

   Let us now consider the linearization of conformal gravity around the
AdS$_6$ background.  For Einstein gravity (\ref{d6einlag}), the
spin-2 graviton is massless, satisfying
\begin{equation}
-(\Box + 2) h_{\mu\nu}^{(m)}=0\,.
\end{equation}
(Recall that we have set $\Lambda=-5$.)
For conformal gravity (\ref{d6conflag}),  the full set of equations
of motion and linearization around the AdS$_6$ vacuum are given in
the appendix.  It turns out the spin-2 modes satisfy
\begin{equation}
\beta (\Box +2)(\Box +6) (\Box
+8)h_{\mu\nu}=0\,.\label{d6conflineareom}
\end{equation}
Thus in the six-dimensional conformal gravity, there are two massive spin-2
modes, with negative
mass-squared, in addition to the massless graviton.  The masses are given by
\begin{equation}
M_1^2=-4\,,\qquad \hbox{and}\qquad M_2^2=-6\,.
\end{equation}
The condition for the absence of tachyon modes is that the lowest energy
$E_0$ of the $SO(2,5)$ representations should be
real, where $E_0$ is given by
\begin{equation}
E_0(E_0 - 5) =M^2\,.
\end{equation}
This implies that
\begin{equation}
M^2 \ge -\fft{25}4\,.\label{d6notachyon}
\end{equation}
Thus both $M_1$ and $M_2$ satisfy this bound, even though both these
massive modes violate the bound $E_0\ge 5$ for unitary representations.
Since they have $M^2<0$, their fall off at large distance is
slower than the modes with
$M^2\ge 0$, and hence they can be truncated out by imposing appropriate
AdS boundary condition whilst the massless graviton is retained.

Use standard Ostrogradsky or Noether techniques, we find that the
on-shell energy of the
massless graviton in the conformal gravity is given by
\begin{equation}
E=\frac{1}{4\kappa^2} (24\beta)\lim_{T\rightarrow\infty}
\frac{1}{T}\int^{T}_{0} dt
\int\sqrt{-g}d^5x\dot{h}_{\mu\nu}\nabla^0h^{\mu\nu}
\end{equation}
For $\beta=-1/24$, this is precisely the on-shell energy of the
Einstein gravity (\ref{d6einlag}), further establishing the
equivalence of Einstein gravity and conformal gravity at the
classical level.

    We may also interpret the above discussion from the point of
view of critical
gravity, whose Lagrangian is given by
\begin{equation}
{\cal L}_6={\cal L}_6^0 - {\cal L}_6^{\rm conf}\,.
\end{equation}
It is clear that the theory admits a unique AdS$_6$ vacuum.
Furthermore, any Einstein metrics with $\Lambda=-5$,
including the Schwarzschild black hole (\ref{d6bh}), are also
solutions. Linearizing the theory around the AdS$_6$ vacuum, it is easy
to verify that the trace mode is trivial and the remaining
spin-2 modes satisfy the equation
\begin{equation}
-(\Box + 2) \Big( 1 +\beta (\Box +6)(\Box+8)\Big)h_{\mu\nu}=0\,.
\end{equation}
Thus in addition to the massless graviton, there are two massive
modes with
\begin{equation}
M_\pm ^2 = -5 \pm \sqrt{1-\fft{1}{\beta}}\,.
\end{equation}
The on-shell energy for the massless graviton is given by
\begin{equation}
E=-\frac{1}{4\kappa^2}(1+24\beta)\lim_{T\rightarrow\infty}
\frac{1}{T}\int^{T}_{0} dt \int\sqrt{-g}d^5x\dot{h}_{\mu\nu}
\nabla^0h^{\mu\nu}\,.\label{d6masslessenergy}
\end{equation}
The criticality condition is
$\beta=-1/24$, for which the massless graviton therefore has
zero energy.  Furthermore,
one of the two massive gravitons becomes massless.  The remaining massive
mode has $M^2=-10$ which violates the no-tachyon bound
(\ref{d6notachyon}).

As in the case of $D=4$, we can consider alternative parameter
choices such that the massive spin-2 modes both satisfy the bound
\begin{equation}
-\fft{25}{4}\le M_\pm^2 <0\,.
\end{equation}
These modes, with non-unitary representations, can nevertheless
be truncated out by imposing appropriate
AdS$_6$ boundary conditions. Furthermore, we would like the
remaining massless graviton to have positive energy, as given by
(\ref{d6masslessenergy}). These requirements can all be met by choosing
\begin{equation}
\beta \ge 1\,.\label{betacon}
\end{equation}
Note that $\beta=1$ corresponds to another critical point, where the two
massive spin-2 modes have the same mass, $M_\pm^2=-5$.  Restoring the general cosmological constant $\Lambda$, defined by $R_{\mu\nu} = \Lambda g_{\mu\nu}$, the condition (\ref{betacon}) becomes $\beta (-\Lambda) \ge 5$.

Finally, we remark that at the $\beta=-1/24$
critical point in the $D=6$ theory,
there are still surviving massive spin-2 modes, since we have only
the one parameter $\beta$ to adjust.  We may also add a Weyl-squared term
$\ft12\alpha \sqrt{-g} C^2$ to the Lagrangian.  The linearized equation
for the spin-2 modes is now given by
\begin{equation}
-(\Box + 2) \Big(1 +\ft32\alpha  (\Box + 6) +\beta (\Box +6)(\Box
+8)\Big)h_{\mu\nu}=0
\end{equation}
A tri-critical point is then achieved with $\alpha=-5/12$ and
$\beta=1/16$, at which the linearized equation becomes
\begin{equation}
-(\Box+2)^3 h_{\mu\nu}=0\,.
\end{equation}

\section{Conclusions}

   In this paper we have developed some new ideas for constructing
higher-derivative theories of gravity that avoid the difficulties with
massive spin-2 ghost modes that typically plague such theories.  In
four dimensions, it was observed in \cite{lpcritical} that if a term
proportional to the square of the Weyl tensor is added to the usual
Einstein-Hilbert Lagrangian with a cosmological constant, then although
generically one finds that the fluctuations around the AdS$_4$ background
describe the usual massless spin-2 graviton and also ghostlike massive
spin-2 modes, it is possible to tune the coefficient of the Weyl-squared
term so as to make the massive modes massless also.  In fact the
energies of the massless modes then vanish in this critical theory.  There
are, however, now also modes with a logarithmic coordinate dependence, which
can have negative energies.  Since their fall off at infinity is slower than
that of the massless modes, they can be removed by imposing appropriate
boundary conditions.  However, since the massless modes that remain have
zero energy, the resulting theory is somewhat trivial.

   Recently, purely conformal gravity where there is only a Weyl-squared
term was revisited \cite{maldaconf}.  In this case there are again
massless and massive spin-2 modes around an AdS$_4$ background, and
again the massive modes are ghostlike.  However, their mass-squared is
actually negative, although not sufficiently negative to be tachyonic.  This
means that they fall off more slowly at infinity than do the massless modes,
and so they can be eliminated by imposing appropriate boundary conditions.
In \cite{maldaconf}, it was shown that by tuning the coefficient of
the Weyl-squared action appropriately, it could be matched for Euclideanised
solutions with the Euclidean cosmological Einstein action for the same
configuration.  It was argued that conformal gravity is then really equivalent
to cosmological Einstein gravity.
In fact the value needed for the Weyl-squared coefficient
is exactly the same as the one required in \cite{lpcritical} for
critical gravity.  This provides a new insight into the trivial
nature of critical gravity once the logarithmic modes are eliminated, in that
its action is essentially just the difference between two actions that
provide equivalent descriptions of long-wavelength physics.

  The main purpose of this paper was to construct a new 1-parameter family
of higher-derivative gravities for which the ghostlike massive spin-2 modes
can be eliminated.  We did this by relaxing the assumption that was made in
\cite{lpcritical} that the mass-squared of the massive spin-2 modes should
be non-negative.  This condition is needed if one wants the massive modes
to carry unitary representations under $SO(2,3)$, but since they are in any
case ghostlike and need to be truncated, this is not really a crucial
requirement.  The key point is that because the background is AdS$_4$ rather
than Minkowski spacetime, there is a window of allowed {\it negative} values
of mass-squared for which the modes are non-tachyonic, and thus classically
stable.  Furthermore, precisely because the mass-squared is negative, the
fall off of these modes is slower than the fall off of the massless modes.
Thus one can impose boundary conditions to eliminate the undesired
massive modes whilst retaining the massless modes.  The massive modes
lie in the desired negative mass-squared range if the parameter $\alpha$
in (\ref{crit0}) satisfies
\be
\alpha \, (-\Lambda) > 12\,,
\ee
i.e. $\alpha>4$ if we normalise the cosmological constant of AdS$_4$
canonically to $\Lambda=-3$.  (Although our primary concern in this paper
is for negative cosmological constant, we expect the above inequality to
hold for positive cosmological constant also.)

   We then extended our discussion to consider gravities in six dimensions.
By taking a suitable linear combination of the three
possible conformally-invariant terms, one can construct a conformal gravity
in six dimensions that admits all Einstein metrics as solutions.  One can
again tune the overall coefficient so that the action of Euclideanised
AdS black holes matches with that calculated
for the cosmological Einstein-Hilbert action.  There are now two sets of
massive spin-2 modes in addition to the massless ones, and both have
mass-squared values that are negative but not tachyonic.
Thus, as in the four-dimensional case studied in \cite{maldaconf}, one
can eliminate the ghostlike massive modes by imposing appropriate
boundary conditions, suggesting the equivalence of Einstein and conformal
gravity in six dimensions too.

   An essential idea underlying the proposal for conformal gravity
in \cite{maldaconf} is that one may be able to ``have one's cake and eat it,''
by reaping the renormalisability benefits of the higher-derivative
theory in the ultraviolet regime, whilst still having an equivalence
to conventional  Einstein gravity in the infrared.
   One motivation for seeking families of potentially acceptable theories
of gravity as we have done in this paper,
rather than isolated examples, comes from quantum considerations.
If one does have a renormalisable theory then the question arises as to
how it behaves in the high-energy limit under the renormalisation group
flow.  One possibility is that the family of theories we have considered
($\alpha\ge 4$ in four dimensions; $\beta\ge1$ in six dimensions) might
start from a finite $\alpha$ or $\beta$ and flow to a fixed point at the
conformally-invariant limit ($\alpha=\infty$ or $\beta=\infty$), as
possibly suggested by the results in \cite{shapiro}.  An advantage of
starting from a finite $\alpha$ or $\beta$ at lower energies, rather than
just using the conformally-invariant theory at all energy scales, would be that
one would in general have the more tightly restricted solution space of
Einstein gravity plus quadratic corrections, flowing to the less restrictive
scale invariance of conformal gravity only in the high-energy limit.
Thus the extensions
of critical gravity we have considered here may be of relevance for a
quantum theory of gravity.

\section*{Acknowledgement}

C.N.P. is grateful to the KITPC, Beijing, for hospitality during the course
of this work.  The research of Y.P. is supported in part by the NSFC under Grant
Nos.10535060/A050207, 10975172 and 10821504, and Ministry of Science
and Technology 973 program under grant No.2007CB815401. The research of
C.N.P. is supported in part by DOE grant DE-FG03-95ER40917.

\appendix

\section{Equations and Linearization of $D=6$ Conformal
Gravity}

\noindent{\bf Equations of motion:}
\medskip

 The Lagrangian for the $D=6$ conformal gravity we study in this
paper is given by (\ref{d6conflag}).  There are five terms.  The
contributions $E_{\mu\nu}^{(i)}$
to Einstein equations of motion from each term is
summarized as follows:
\begin{eqnarray}
1):&&RR^{\mu\nu}R_{\mu\nu}\Rightarrow \cr 
&&E^{(1)}_{\mu\nu}=
 (\Box (R_{\lambda\sigma}R^{\lambda\sigma})+\nabla_{\lambda}\nabla_{\sigma}
(RR^{\lambda\sigma}) -\ft{1}{2}RR_{\lambda\sigma}R^{\lambda\sigma})g_{\mu\nu}+R_{\lambda\sigma}R^{\lambda\sigma}R_{\mu\nu} +
2RR_{\lambda\mu} R^{\lambda}_{~\nu}\cr 
&&\qquad\quad+\Box(RR_{\mu\nu})-\nabla_{\mu}\nabla_{\nu}(R_{\lambda\sigma}R^{\lambda\sigma}) -
\nabla_{\lambda}\nabla_{\mu}(RR^{\lambda}_{~\nu})
-\nabla_{\lambda}\nabla_{\nu}(RR^{\lambda}_{~\mu})\,,\cr 
2):&&R^3\Rightarrow\cr 
&&E^{(2)}_{\mu\nu}=(3\Box R^2-\ft{1}{2}R^3)g_{\mu\nu}+3R^2R_{\mu\nu} -
3\nabla_{\mu}\nabla_{\nu}R^2\,,\cr 
3):&&R^{\mu\nu}R^{\lambda\rho}R_{\mu\lambda\nu\rho}
\Rightarrow\cr 
&&E^{(3)}_{\mu\nu}=-\ft{1}{2}R^{\sigma\delta}R^{\lambda\rho}
R_{\sigma\lambda\delta\rho}g_{\mu\nu} + \ft{3}{2}R^{\rho\sigma}
R_{\rho\mu\sigma\lambda} R^{\lambda}_{~\nu}
+\ft{3}{2}R^{\rho\sigma} R_{\rho\nu\sigma\lambda}
R^{\lambda}_{~\mu}\cr 
&&\qquad\quad +\Box(R^{\rho\sigma}R_{\rho\mu\sigma\nu}) +
\nabla^{\sigma}\nabla^{\delta} (R^{\lambda\rho}
R_{\lambda\sigma\rho\delta})g_{\mu\nu}\cr 
&&\qquad\quad -\nabla^{\lambda}\nabla_{\mu} (R^{\rho\sigma}
R_{\rho\lambda\sigma\nu})-\nabla^{\lambda}\nabla_{\nu} (R^{\rho\sigma}
R_{\rho\lambda\sigma\mu})\cr 
&&\qquad\quad -\nabla_{(\sigma}\nabla_{\lambda)}
(R_{\mu}^{~\sigma}R_{\nu}^{~\lambda}) +
\nabla_{\sigma}\nabla_{\lambda}(R_{\mu\nu}R^{\sigma\lambda})\cr 
4):&&R^{\mu\nu}\Box R_{\mu\nu} = -g^{\mu\nu}
\nabla_{\mu} R^{\lambda\rho}\nabla_{\nu}R_{\lambda\rho}
\Rightarrow\cr 
&&E^{(4)}_{\mu\nu}= \ft{1}{2} g_{\mu\nu}(g^{\sigma\delta} \nabla_{\sigma}
R^{\lambda\rho} \nabla_{\delta}R_{\lambda\rho}) -(2\nabla^{\sigma}
R_{\mu\lambda}\nabla_{\sigma} R_{\nu}^{~\lambda} +
\nabla_{\mu}R_{\sigma\lambda} \nabla_{\nu}R^{\sigma\lambda})\cr 
&&\qquad\quad +2\nabla_{\lambda} (R_{\sigma(\mu}\nabla_{\nu)}
R^{\lambda\sigma}) +2\nabla_{\lambda} (\nabla^{\lambda}
R^{\sigma}_{~(\nu}R_{\mu)\sigma}) -2\nabla_{\sigma}(\nabla_{(\mu}
R_{\nu)\lambda}R^{\sigma\lambda})\cr 
&&\qquad\quad + \Box^2R_{\mu\nu}+\nabla_{\sigma}\nabla_{\lambda}\Box
R^{\sigma\lambda}g_{\mu\nu}-\nabla_{\lambda}\nabla_{\nu}(\Box
R^{\lambda}_{~\mu})-\nabla_{\lambda}\nabla_{\mu}(\Box
R^{\lambda}_{~\nu})\cr 
5):&&R\Box R=-g^{\mu\nu}\nabla_{\mu}R\nabla_{\nu}R
\Rightarrow \cr 
&&E^{(5)}_{\mu\nu}= \ft{1}{2}g_{\mu\nu}
(g^{\sigma\lambda}\nabla_{\sigma}R\nabla_{\lambda}R)
-\nabla_{\mu}R\nabla_{\nu}R+2(\Box R)R_{\mu\nu}\cr 
&&\qquad\quad +2(\Box^2R)g_{\mu\nu}-2\nabla_{\mu}\nabla_{\nu}\Box R
\end{eqnarray}
The complete equation of motion following from (\ref{d6conflag}) is then
\be
E^{(1)}_{\mu\nu} - \ft3{25} E^{(2)}_{\mu\nu} - 2 E^{(3)}_{\mu\nu}
 - E^{(4)}_{\mu\nu} + \ft3{10} E^{(5)}_{\mu\nu}=0\,.
\ee

\bigskip
\noindent{\bf Linearization:}
\medskip

The theory admit Einstein metrics with $R_{\mu\nu}=\Lambda
g_{\mu\nu}$, with arbitrary $\Lambda$.  We consider the
linearization around the AdS$_6$ background, namely
$R_{\mu\nu\rho\sigma}= \fft15\Lambda (g_{\mu\rho} g_{\nu\sigma} -
g_{\mu\sigma} g_{\nu\rho})$.  Writing the varied metric as
$g_{\mu\nu}\rightarrow g_{\mu\nu} + h_{\mu\nu}$, and so $\delta
g_{\mu\nu}=h_{\mu\nu}$, the linearized Einstein tensor is given by
\begin{eqnarray}
\cG_{\mu\nu}^L &=& R^L_{\mu\nu} -\ft12 R^L\, g_{\mu\nu} -\Lambda\,
h_{\mu\nu} \,,\qquad \cG^L\equiv g^{\mu\nu}
\cG^L_{\mu\nu}\,,\label{GL}\\ 
R^L_{\mu\nu} &=& \nabla^\lambda\nabla_{(\mu} h_{\nu)\,\lambda}
-\ft12\square h_{\mu\nu} -\ft12 \nabla_\mu\nabla_\nu h\,,\label{RicL}\\
R^L&=& \nabla^\mu\nabla^\nu h_{\mu\nu} -\square h - \Lambda
h\,.\label{RL}
\end{eqnarray}
(We have also defined $R^L_{\mu\nu}$, the linearization of
$R_{\mu\nu}$, and introduced $h=g^{\mu\nu} h_{\mu\nu}$.) With these
preliminaries, we find that the linearized five contributions of the
equations of motion listed above are given by
\begin{eqnarray}
1): &&(2\Lambda\Box \cG^L+ 6\Lambda\nabla^{\lambda}\nabla^{\sigma}
\cG^L_{\lambda\sigma}-4\Lambda^2\cG^L)g_{\mu\nu} +30\Lambda^2
\cG^L_{\mu\nu}+6\Lambda\Box \cG^L_{\mu\nu}\cr 
&&-2\Lambda\nabla_{\mu}\nabla_{\nu}\cG^L -
6\Lambda\nabla^{\sigma}\nabla_{\mu}\cG^L_{\sigma\nu} -
6\Lambda\nabla^{\sigma}\nabla_{\nu}\cG^L_{\sigma\mu}+14\Lambda
g_{\mu\nu}\Box R^L \cr 
&& +2\Lambda^2g_{\mu\nu}R^L
-14\Lambda\nabla_{\mu}\nabla_{\nu}R^L\,,\cr 
2):&& 108\Lambda^2\cG^L_{\mu\nu} + 36\Lambda g_{\mu\nu}\Box R^L +
 36\Lambda^2 g_{\mu\nu}R^L-36\nabla_{\mu}\nabla_{\nu}R^L\,,\cr 
3): && 3\Lambda R_{\mu~\nu}^{~\lambda~\sigma}
\cG_{\lambda\sigma}^L+6\Lambda^2
\cG_{\mu\nu}^L+\Lambda\Box\cG^L_{\mu\nu} +\Box
(R_{\mu~\nu}^{~\lambda~\sigma}\cG_{\lambda\sigma}^L)
+g_{\mu\nu}\nabla^{\sigma}\nabla^{\delta}
(R_{\sigma~\delta}^{~\lambda~\rho}\cG^L_{\lambda\rho})\cr 
&& -\nabla^{\lambda}\nabla_{\mu} (R_{\lambda~\nu}^{~\sigma~\delta}
\cG^L_{\sigma\delta}) -\nabla^{\lambda}\nabla_{\nu}
(R_{\lambda~\mu}^{~\sigma~\delta}\cG^L_{\sigma\delta})
-\ft{3}{2}\Lambda\nabla^{\sigma}\nabla_{\mu} \cG^L_{\sigma\nu} -
\ft{3}{2}\Lambda\nabla^{\sigma}\nabla_{\nu}\cG^L_{\sigma\mu}\cr 
&&+\Lambda\Box \cG^L_{\mu\nu}+3\Lambda^2g_{\mu\nu}R^L+3\Lambda
g_{\mu\nu}\Box R^L-3\Lambda\nabla_{\mu}\nabla_{\nu}R^L\,,\cr 
4):&& 2\Lambda\Box\cG^L_{\mu\nu} + \Box^2\cG^L_{\mu\nu} +
g_{\mu\nu}\nabla^{\lambda}\nabla^{\sigma} \Box\cG^L_{\lambda\sigma}
-\nabla^{\lambda}\nabla_{\mu} \Box\cG^L_{\lambda\nu}
-\nabla^{\lambda}\nabla_{\nu}\Box\cG^L_{\lambda\mu} \cr 
&&+g_{\mu\nu}\Lambda\Box R^L + g_{\mu\nu}\Box^2R^L -
\nabla_{\mu}\nabla_{\nu}\Box R^L\,,\cr 
5):&& 2(\Lambda g_{\mu\nu}\Box R^L +
g_{\mu\nu}\Box^2R^L-\nabla_{\mu}\nabla_{\nu}\Box R^L)\,.
\end{eqnarray}
Thus for the traceless and transverse spin-2 modes $h_{\mu\nu}$, the
linearized Einstein tensor is $\cG_{\mu\nu}^L= -\ft12 (\Box + 2) h_{\mu\nu}$
and the Ricci-scalar is $R^{L}=0$.  It follows that the linearized
equation of motion is given by (\ref{d6conflineareom}).

\bigskip
\noindent{\bf Hamiltonian:}
\medskip

The quadratic fluctuations $S_2$ for the following action $S$ are given
by
\begin{equation}
S=\ft{1}{2\kappa^2}\int\sqrt{-g}d^6x \bigl[R-4\Lambda-\beta(RR^{\mu\nu}
R_{\mu\nu}-\ft{3}{25}R^3 -
2R^{\mu\nu}R^{\lambda\rho}R_{\mu\lambda\nu\rho} -R^{\mu\nu}\Box
R_{\mu\nu}+\ft{3}{10}R\Box R)\bigr]\,,
\end{equation}
\begin{eqnarray}
S_2=&&-\frac{1}{8\kappa^2}\int\sqrt{-g}d^6x \biggl[\nabla_{\lambda}
h_{\mu\nu}\nabla^{\lambda}
h^{\mu\nu}-2h_{\mu\nu}h^{\mu\nu}+\beta(\nabla_{\lambda} \Box
h_{\mu\nu}\nabla^{\lambda}\Box h^{\mu\nu}-16\Box
h^{\mu\nu}\Box h_{\mu\nu}\cr 
&&+76\nabla_{\lambda} h_{\mu\nu}\nabla^{\lambda}
h^{\mu\nu}-96h^{\mu\nu}h_{\mu\nu})\biggr]\,.
\end{eqnarray}
The Hamiltonian is
\begin{equation}
H=\lim_{T\rightarrow\infty}\frac{1}{T}\int^{T}_{0}dt
\int d^5x \Big(\dot{h}_{\mu\nu}\Pi^{(1)\mu\nu}
+\partial_t(\nabla_0h_{\mu\nu})\Pi^{(2)\mu\nu}
+\partial_t(\Box h_{\mu\nu})\Pi^{(3)\mu\nu}-L\Big)\,.
\end{equation}
where
\begin{eqnarray}
\Pi^{(1)\mu\nu}&=&-\frac{\sqrt{-g}}{4\kappa^2}
[\nabla^0h^{\mu\nu}+\beta(76\nabla^0h^{\mu\nu} +
16\nabla^0\Box h^{\mu\nu}+\nabla^0\Box^2h^{\mu\nu})]\cr 
\Pi^{(2)\mu\nu} &=&-\frac{\beta\sqrt{-g}}{4\kappa^2}
[-16g^{00}\Box h^{\mu\nu}-g^{00}\Box^2h^{\mu\nu}]\cr 
\Pi^{(3)\mu\nu} &=&-\frac{\beta\sqrt{-g}}{4\kappa^2}[
\nabla^0\Box h^{\mu\nu} ]
\end{eqnarray}
Then one can obtain the energy of massless graviton as
\begin{equation}
E=-\frac{1}{4\kappa^2}(1+24\beta) \lim_{T\rightarrow\infty}
\frac{1}{T} \int^{T}_{0} dt\int\sqrt{-g}
d^5x\dot{h}_{\mu\nu}\nabla^0h^{\mu\nu}
\end{equation}
In our case, the Wald formula is
\begin{equation}
S=-\frac{1}{8G}\int_{H} \epsilon_{ab}\epsilon_{cd}\Big(\frac{\partial
L}{\partial R_{abcd}}+\nabla_{(mn)}\frac{\partial
L}{\partial\nabla_{(mn)} R_{abcd}}\Big)d\Sigma,
\end{equation}
where $\epsilon_{ab}$ is the bi-normal vector of horizon normalized
to satisfy $\epsilon_{ab}\epsilon^{ab}=-2$.

\end{document}